\documentclass[aps,prb,twocolumn,showpacs,bibnotes,reprint,superscriptaddress]{revtex4-1}

\pdfoutput=1

\usepackage{amsfonts}
\usepackage{amsmath}
\usepackage{bm}
\usepackage{bbm}
\usepackage{braket}
\usepackage{mhchem}
\usepackage{paralist}
\usepackage{url}
\usepackage{todonotes}
\usepackage{subcaption}
\usepackage[capitalise]{cleveref}
\crefname{appsec}{Appendix}{Appendices}

\newcommand{\HF}{\ensuremath{\textrm{HF}}}
\newcommand{\pclust}{\ensuremath{p_{\textrm{cluster}}}}
\newcommand{\psize}{\ensuremath{p_{\textrm{size}}}}
\newcommand{\Ntot}{\ensuremath{N_{\textrm{total}}}}
\newcommand{\Nmove}{\ensuremath{{\nu_{\textrm{move}}}}}
\newcommand{\rs}{\ensuremath{r_{\textrm{s}}}}
\newcommand{\bohr}{\ensuremath{\mathrm{a_0}}}
\newcommand{\hartree}{\ensuremath{\mathrm{E_h}}}
\newcommand{\Np}{\ensuremath{N_{\mathrm{p}}}}
\DeclareMathOperator{\sgn}{sgn}
\DeclareMathOperator{\hash}{hash}

\begin{document}

\title{Large Scale Parallelization in Stochastic Coupled Cluster}

\author{J.~S.~Spencer}
\affiliation{Department of Materials, Imperial College London, Exhibition Road, London, SW7 2AZ, U.K.}
\affiliation{Department of Physics, Imperial College London, Exhibition Road, London, SW7 2AZ, U.K.}
\author{V.~A.~Neufeld}
\affiliation{University Chemical Laboratory, Lensfield Road, Cambridge, CB2 1EW, U.K.}
\author{W.~A.~Vigor}
\affiliation{Department of Chemistry, Imperial College London, Exhibition Road, London, SW7 2AZ, U.K.}
\author{R.~S.~T.~Franklin}
\author{A.~J.~W.~Thom}
\email{ajwt3@cam.ac.uk}
\affiliation{University Chemical Laboratory, Lensfield Road, Cambridge, CB2 1EW, U.K.}

\begin{abstract}
    Coupled cluster theory is a vital cornerstone of electronic structure theory and is being applied to ever-larger systems.
    Stochastic approaches to quantum chemistry have grown in importance and offer compelling advantages over traditional deterministic algorithms in terms of computational demands, theoretical flexibility or lower scaling with system size.
    We present a highly parallelizable algorithm of the coupled cluster Monte Carlo method involving sampling of clusters of excitors over multiple time steps.
    The behaviour of the algorithm is investigated on the uniform electron gas and the water dimer at CCSD, CCSDT and CCSDTQ levels.
    We also describe two improvements to the original sampling algorithm, \textit{full non-composite} and
     \textit{multi-spawn}
    sampling.
    A stochastic approach to coupled cluster results in an efficient and scalable implementation at arbitrary truncation levels in the coupled cluster expansion.
\end{abstract}

\date{\today}
\maketitle

\section{Introduction}
\label{sec:intro}

Coupled cluster (CC) methods\cite{BartlettMusial_07RMP} are of crucial importance in electronic structure and have been used to explore a variety of systems, including atoms and molecular systems\cite{BartlettMusial_07RMP,KowalskiPiecuch_00JCP,HardingStanton_08JCP,HattigTew_12CR,Karton_16CMS}, the uniform electron gas\cite{Freeman1977,Bishop1978, Bishop1982,Shepherd2013, Roggero2013, Spencer2016, McClain2016a, Shepherd2016a, Neufeld2017} and solids/other periodic systems\cite{Hirata2001a,Hirata2004,Manby2006, Nolan2009, Gruneis2011, Booth2013, Gruneis2015, Liao2016, Schwerdtfeger2016, McClain2017, Gruber2018}.
CCSD(T)\cite{Raghavachari1989}, where single and double excitations are included in the wavefunction ansatz and supplemented with the pertubative treatment of triple excitations, is commonly regarded as the ``gold standard'' of quantum chemistry and can frequently achieve\cite{Lee1995} chemical accuracy of 1 kcal/mol.

Despite these successes, coupled cluster is not without its drawbacks.
Coupled cluster is systematically improvable, at least in principle, by increasing the excitation level included in the CC wavefunction ansatz.
Doing so makes the conventional CC equations vastly more complicated and hence computational demanding.
As a result, treating higher truncation levels is possibly only in specialist codes\cite{MRCC}.
Conventional implementations of coupled cluster also rely heavily upon dense linear algebra, which does not scale well with increasing numbers of processors on parallel or heterogeneous computer architectures, though recent work in linear algebra and tensor libraries are making impressive progress\cite{Agullo2009,Solomonik2014}.

One avenue for improving the computational efficiency of coupled cluster is to exploit the nearsighted nature of electron correlation and use local approximations\cite{FlockeBartlett_04JCP,Ziokowski2010,RiplingerNeese_13JCP,RiplingerNeese_13JCP2}.
Another approach, of increasing use in quantum chemistry and the broader electronic structure community, is to use stochastic methods; these have proven to provide low-scaling algorithms for electronic structure methods and typically exhibit excellent scaling with increasing processor count\cite{Foulkes2001,Thom2007,Willow2012,Willow2013,Baer2013}.
Local and stochastic methods may also be easily combined via a localisation transformation of the mean-field single-particle orbitals.\footnote{Our implementation does not require canonical orbitals, so we have the full freedom to choose
any transformation of the single particle orbitals, for example the study in Ref. \onlinecite{Neufeld2018}}.

The full configuration interaction quantum Monte Carlo (FCIQMC) method\cite{BoothAlavi_09JCP,ClelandAlavi_12JCTC} has been a major development in quantum chemistry. By sampling the action of the Hamiltonian, FCIQMC has been able to calculate exact properties for quantum systems inaccessible to conventional diagonalisation techniques\cite{BoothAlavi_09JCP, ShepherdAlavi_12PRB, Overy2014, Blunt2015, Blunt2017}.  The computational advantage of FCIQMC is largely through a representation of the FCI wavefunction which is significantly more compact than the full wavefunction, though still scaling factorially with the size of the Hilbert space sampled.

One of us (AJWT) subsequently used a similar approach to formulate a Monte Carlo approach to coupled cluster theory (CCMC)\cite{Thom_10PRL}, inheriting the benefits of more compact storage, and now scaling with the polynomial size of the truncated CC space.
The initiator approximation can substantially improve the stochastic sampling of the wavefunction in both FCIQMC\cite{ClelandAlavi_12JCTC} and CCMC\cite{Spencer2016}, though the latter requires careful extrapolation.
The stochastic sampling of the coupled cluster wavefunction can be further improved by sampling only linked diagrams\cite{FranklinThom_16JCP} and non-uniform sampling of the coupled cluster expansion \cite{Scott2017}, and improved sampling of the action of the Hamiltonian\cite{Neufeld2018}.
The utility of CCMC has been demonstrated to calculate coupled cluster energies at up to the CCSDTQ56 level for molecular systems\cite{Thom_10PRL,Scott2017}, for the uniform electron gas\cite{Spencer2016,Neufeld2017} and also been used to automatically generate the $P$ subspace in the CC($P;Q$) method\cite{Deustua2017}.
However, due to the non-linearity of the coupled cluster equations, parallelization of the CCMC algorithm is less straightforward than a parallel FCIQMC implementation\cite{Booth2014}.

We present a brief overview of the CCMC algorithm in \cref{sec:CCMC} provide context to the problem. In \cref{sec:CCMC_MPI} we show that the CCMC algorithm can be efficiently parallelized by introducing an additional level of Monte Carlo sampling by considering only a subset of terms in the coupled cluster expansion per iteration. The accuracy and performance of this algorithm is investigated using the uniform electron gas and the water dimer. \cref{sec:sampling} provides simple improvements to the original CCMC algorithm to improve stability and convergence of the CC wavefunction.
We conclude in \cref{sec:discussion}.

\section{Coupled Cluster Monte Carlo}
\label{sec:CCMC}

The algorithms used to sample the FCI and coupled cluster wavefunctions have been previously detailed\cite{BoothAlavi_09JCP,Spencer2012,Thom_10PRL,Spencer2016,Scott2017} and as such we summarise the key features relevant to this work here.

The coupled cluster wavefunction ansatz can be expressed as $\ket{\Psi} = N e^{\hat{T}} \ket{D_\HF}$, where $\ket{D_\HF}$ is the Hartree--Fock determinant, $N$ controls the (intermediate) normalisation and the cluster operator $\hat{T}$ is
\begin{equation}
    \hat{T} = \sum_{i,a} t_{i}^{a} \hat{c}_{i}^{a} + \sum_{\substack{i<j\\a<b}} t_{ij}^{ab} \hat{c}_{ij}^{ab} + \cdots,
\end{equation}
where $\{t_{i\cdots}^{a\cdots}\}$ is the set of amplitudes and $\hat{c}_{i\cdots}^{a\cdots}$ is an \emph{excitor} comprising of a string of creation and annihilation operators.  For convenience, we use $\hat{c}_{\bm{i}}$ and $t_{\bm{i}}$, such that $\hat{c}_{\bm{i}}\ket{D_\HF}$ produces $\ket{D_{\bm{i}}}$ (up to a sign, as discussed later) and $t_{\bm{i}}$ is the corresponding amplitude, and rescale the amplitudes with an additional factor, $t_{\HF}$, such that 
\begin{equation}
    \ket{\Psi} = t_{\HF} e^{\hat{T}/t_{\HF}} \ket{D_\HF}.
\end{equation}
Within this wavefunction ansatz, the coefficient of a given determinant, $\tilde{t}_{\bm{j}}=\braket{D_{\bm{j}}|\Psi}$, contains contributions from all sets of excitors which can be combined to produce that determinant.

As with FCIQMC, CCMC applies an approximate linear propagator, $1-\delta\tau (\hat H-S)$, where $\delta\tau$ is the timestep and $S$ is an adjustable parameter to control proportionality, and which has the same eigenspectrum as $e^{-\delta\tau \hat H}$ for sufficiently small $\delta\tau$\cite{Spencer2012}. Applying this to $\ket{\Psi}$ and cancelling quadratic and higher-order terms\cite{Thom_10PRL, Spencer2016} results in a form reminiscent of the propagation equation for FCIQMC\cite{BoothAlavi_09JCP},
\begin{equation}
    t_{\bm{i}}(\tau+\delta\tau)=t_{\bm{i}}(\tau) - \delta\tau \sum_{\bm{j}} (H_{\bm{ij}} - S\delta_{\bm{ij}}) \tilde{t}_{\bm{j}}(\tau).
    \label{eqn:CCMC}
\end{equation}
The similarity-transformed Hamiltonian can also be used in the projection the coupled cluster wavefunction\cite{FranklinThom_16JCP} and the approaches presented here equally apply to that formulation.
The amplitudes are stochastically sampled by representing them using either particles with integer\cite{BoothAlavi_09JCP,Thom_10PRL} or real (as opposed to integer) weights\cite{Petruzielo2012,Overy2014}, which has been shown to reduce stochastic error within FCIQMC and readily applies to CCMC.

In FCIQMC, the particles on each occupied determinant are explicitly evolved; in CCMC this would require one to first evaluate all possible $\tilde{t}_{\bm{j}}$, which is computationally painful. Instead, we exploit the fact that Monte Carlo is a powerful tool for sampling high dimensional spaces and, in addition to stochastically sampling the action of the Hamiltonian, also sample the wavefunction ansatz. The algorithm used to sample the cluster expansion has been shown to have a significant impact on computational and statistical efficiency\cite{Scott2017}; here we consider only the simplest approach.  
The cluster size, $s=[0, l+2]$, is selected according to an exponential distribution, i.e. $\psize(s) = 2^{-(s+1)}$, where $l$ is the highest order term in the cluster expansion\footnote{We set $\psize(s) = 1 - \sum_{i=0}^{s-1} \psize(i)$ in order to ensure a normalised probability.}. The cluster is then generated by selecting $s$ excitors from the current distribution, each with probability $|t_{\bm{i}}| / (\Ntot-t_{\HF})$, where $\Ntot$ is the total current population. A cluster containing the same excitor more than once is discarded.
Alternative approaches for sampling the cluster expansion are discussed in \cref{sec:sampling}.

The dynamics for evolving the particles on a cluster are essentially identical to those in FCIQMC\cite{BoothAlavi_09JCP}.
A simulation starts with a number of particles of unit weight on the Hartree--Fock determinant. The particles are then evolved by sampling the action of the Hamiltonian on each particle, allowing new particles to be created (`spawned'), and the particle to die (due to the sign of the Hamiltonian operator). At the end of each iteration particles on the same excitor with opposite signs are removed (`annihilated') from the simulation, which aids the sign problem\cite{Spencer2012} and is a statistically exact process. Note that for clusters of size 2 and higher, the death step amounts to creating a particle of opposite sign on the corresponding excitor.
Events which create particles on excitors which are not within the desired truncation level of the cluster operator are simply discarded in our current CCMC implementation.

Anti-commutation relationships in strings of creation and annihilation operators must be handled with care.  A given excitor is required to be unique and hence an arbitrary excitor $\{\hat{c}_{ij\cdots l}^{ab\cdots e}\}$ must satisfy $i<j<\cdots<l$ and $a<b<\cdots<e$.  Defining $\hat{c}_i^{\vphantom{}}$ ($\hat{c}_i^\dagger$) to annihilate (create) an electron in the $i$-th spin-orbital, an excitor and a determinant can be expressed as
\begin{gather}
    \hat{c}_{ij\cdots l}^{ab\cdots e} = \hat{c}_a^\dagger \hat{c}_b^\dagger \cdots \hat{c}_e^\dagger \hat{c}_l^{\vphantom{}} \cdots \hat{c}_j^{\vphantom{}} \hat{c}_i^{\vphantom{}} \\
    \ket{D_{\bm{i}}} = \ket{i_1 i_2 i_3 \cdots i_N} = \hat{c}_{i_1}^\dagger \hat{c}_{i_2}^\dagger \hat{c}_{i_3}^\dagger \cdots \hat{c}_{i_N}^\dagger \ket{0},
\end{gather}
where $\ket{0}$ is the vacuum state and $i_1<i_2<\cdots<i_N$.
Therefore, when collapsing a cluster, $\hat{c}_{\bm{i}}\hat{c}_{\bm{j}}\ldots\hat{c}_{\bm{k}}$, to a single excitor, $\hat{c}_{\bm{l}}$, a negative sign must be included as required by anticommutativity in order for the operators in the cluster to match the order in the single excitor.
Similarly when an excitor is applied to the Hartree--Fock determinant, the resultant set of creation operators must be permuted in order to achieve the required ordering:
\begin{equation}
    s_D(\bm{i}) = \braket{D_{\bm{i}} | \hat{c}_{\bm{i}} | D_\HF}.
\end{equation}
The sign from collapsing a cluster is conveniently absorbed into the amplitude of the cluster and the sign from converting to/from a determinant in the spawning step, such that the sign of the spawned particle is determined by $-\sgn\left(H_{\bm{i}\bm{j}}s_D(\bm{i})s_D(\bm{j})\right)$.

The energy shift, $S$, is not known \emph{a priori}. In keeping with other QMC methods\cite{Umrigar1993,BoothAlavi_09JCP}, $S$ is updated to keep the population stable. 
In a simulation $S$ is initially held constant (typically at the Hartree--Fock energy) to allow the population to grow and is only adjusted once the population has reached a desired value. It is important to take the non-linear wavefunction ansatz into account during the constant-shift phase in order to ensure correct normalisation\cite{FranklinThom_16JCP}.

The energy at a given time can, as with FCIQMC, be evaluated with a projected estimator.
Again, it is simpler to sample the the wavefunction using the same set of clusters, $\{\hat{c}_{\bm{j}}\}$, chosen above:
\begin{equation}
    E_{\textrm{proj.}} 
                        = t_{\HF}^{-1} \sum_{\{\hat{c}_{\bm{j}}\}} \braket{D_\HF|\hat{H}|D_{\bm{j}}} \frac{t_{\bm{j}} s_D(\bm{j})}{\pclust(\bm{j})}.
\end{equation}

\section*{Computational Methods}

All CCMC calculations are performed using a development version of HANDE\cite{HANDEpaper,Spencer2018c}. Most one- and
two-body molecular integrals were obtained from restricted Hartree--Fock calculations performed in Psi4\cite{Psi4},
except for the study of three water molecules at large distances from each other where the integrals were
obtained with PySCF \cite{Sun2018} and localised with a Boys method\cite{Foster1960} by PySCF.
Floating-point weights were used to improve stochastic efficiency.
Input files and raw data are available under a Creative Commons license at \url{https://doi.org/10.17863/CAM.30359}.
Estimates of the stochastic error in CCMC simulations were obtained via a reblocking analysis\cite{Flyvbjerg1989}.
QMC energies were verified to be unaffected by a population control bias by comparison to 
those obtained using a reweighting analysis\cite{Umrigar1993,Vigor2015}.

All data was analysed using numpy\cite{Oliphant2015}, pandas\cite{Mckinney2010} and pyblock\cite{pyblock} and plots produced using matplotlib\cite{Hunter2007} and seaborn\cite{seaborn}.

\section{Parallelisation}
\label{sec:CCMC_MPI}

Using distributed computer architectures is advantageous both in terms of reducing the runtime of a calculation and in being able to treat larger systems due to the corresponding increase in available memory.
The memory usage on a given processor of a QMC calculation in Slater determinant space is proportional to the number of states stored on that processor whilst the computational workload is a function of both the number of states and the total population on the processor\cite{Booth2014}.
Ideally both would be evenly balanced across all processors whilst the annihilation step requires all particles on the same determinant to be placed on the same processor.
The size of the Hilbert space precludes a lookup table and simply dividing the Hilbert space into chunks and assigning chunk(s) to a processor yields poor load balancing as the distribution of `important' states tends to be highly irregular in many chemical systems.  Booth \emph{et al.}\cite{Booth2014} proposed a deterministic mapping of a determinant to a processor, $p(\ket{D})$, in a time- and space-efficient manner:
\begin{equation}
    p(\ket{D}) = \hash(\ket{D}) \bmod \Np
    \label{eq:fciqmc_dist}
\end{equation}
where $\Np$ is the number of processors and $\hash$ is a function which maps an arbitrary amount of data (here a representation of a determinant) to an integer over a fixed range.  Crucially a good hash function returns different values for similar inputs and hence determinants which are close in excitation space are mapped to different processors.\footnote{Booth \emph{et al.} used a custom hash function based upon the list of occupied orbitals.  We find hashing the bit string representation of the determinant simpler and computationally more efficient, whilst giving at least as good distribution over processors when a hash function of sufficient quality is used, and therefore use the MurmurHash2 function.}

CCMC introduces the additional complication that the cluster expansion must be sampled. One option, which we exploit, is to use a shared-memory paradigm (implemented in HANDE using OpenMP), where the cluster selection and evolution are distributed over threads.
Distributing the set of states over multiple nodes\footnote{A node may consist of a single processor or multiple processors. Within the MPI paradigm, we distribute over MPI ranks, where each rank contains one or more threads.}, as done in FCIQMC, is not helpful as either each spawning event would involve communication between nodes in order to randomly generate clusters. Instead, we again exploit Monte Carlo sampling: a node only samples the subset of clusters that can be formed from the excitors residing on that node. Crucially, the subset of clusters changes such that all clusters have an equal chance of being selected within a few timesteps.

Concretely, \cref{eq:fciqmc_dist} is extended to periodically change the processor of a given excitor:
\begin{gather}
    o(\ket{D},i_\tau) = (\hash(\ket{D}) + i_\tau) \gg \Nmove \\
    p(\ket{D},i_\tau) = \hash\left(\ket{D} \oplus o(\ket{D},i_\tau)\right) \bmod \Np
    \label{eq:ccmc_dist}
\end{gather}
where $i_\tau$ is the iteration index at time $\tau$ (i.e. $\tau/\delta\tau$), $2^{\Nmove}$ is a constant termed the `move frequency' and is discussed below, $a \gg b$ represents the right-shift bit operation on $a$, where the bits in $a$ are moved to the right and the $b$ least significant bits are removed,  and $\oplus$ is the exclusive or bit operation.
The offset function, $o(\ket{D}, i_\tau)$, discards the lower $\Nmove$ bits and hence changes value every $2^{\Nmove}$ iterations, where the iteration at which it first changes is determined by the value of $\hash(\ket{D})$. 
Hence the processor index of an excitor can change every $2^{\Nmove}$ iterations. Given a good hash function, $p$ returns a even distribution of values in the range $[0,\Np)$ and all clusters can still be sampled over a number of iterations. The probability of selecting a cluster of size $s$ is scaled by $\Np^{-s}$ to account for the probability of each excitor in the cluster being on the same processor.
Excitors are efficiently redistributed at the same time as newly created particles are communicated to the appropriate node.

\subsection{Parallelization Scaling}
\begin{figure}
    \centering
    \includegraphics[width=1.0\linewidth,keepaspectratio]{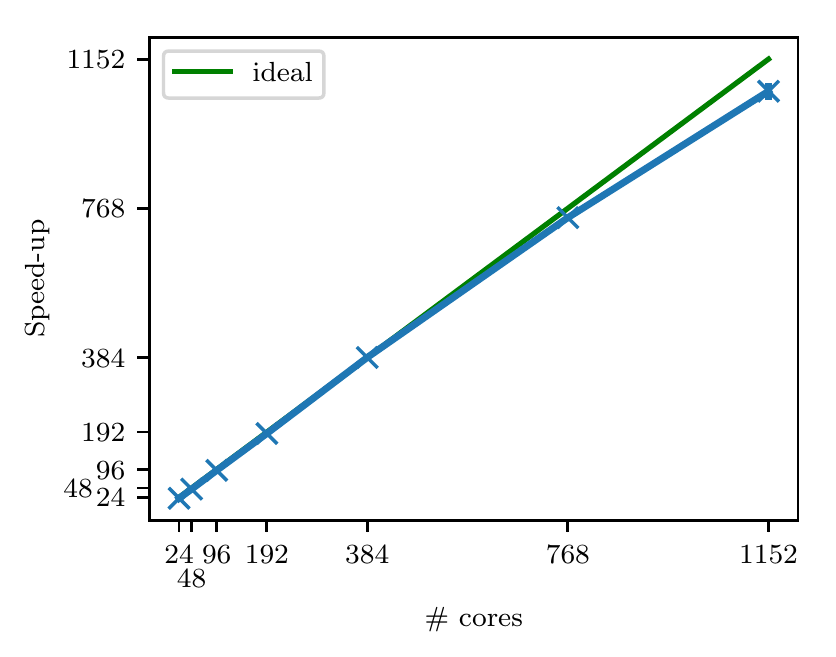}
    \caption{Scaling of hybrid MPI+OpenMP CCSDT calculations on the water dimer using a jun-cc-pVDZ basis set performed using even selection\cite{Scott2017}. 12 OpenMP threads were used per MPI process. Timings were taken from an equilibrated calculation on 384 cores restarted on different numbers of cores.  Error bars are only visible for 1152 cores. }
    \label{fig:scaling}
\end{figure}

The scaling of the parallelization algorithm is demonstrated on the water dimer in \cref{fig:scaling}. All calculations were run with a different number of MPI processes divided into 12 OpenMP threads giving a total number of cores used. They were all restarted from a calculation that was run on 384 cores (32 MPI processes $\times$ 12 OpenMP threads) and used the same parameters. The speed-up was then evaluated as the ratio of time taken per iteration when using 384 cores over the current number of cores.
Up to about 500 cores, the `strong scaling' is approximately ideal. After 1000 cores, over 90\% of ideal scaling is still achieved. The calculations used about 1.5$\times 10^7$ excips and had about 8$\times 10^6$ occupied states/excitors. The scaling depends upon the effect of load-balancing, and the ratio of calculation to communication time, both of which reduce efficiency as the number of occupied excitors per core decreases.  For calculations with over $10^4$ excitors per core we find no loss of computational efficiency upon parallelization. As system size increases, the number of excitors grows polymonially, so in this `weak scaling' regime the algorithm displays perfect parallelization, over 1000 cores can be employed for sufficient calculation size.

\subsection{Parallelization Bias}
\begin{figure*}
\centering
	\begin{subfigure}[h]{\linewidth}
    \includegraphics[width=1.0\linewidth,keepaspectratio]{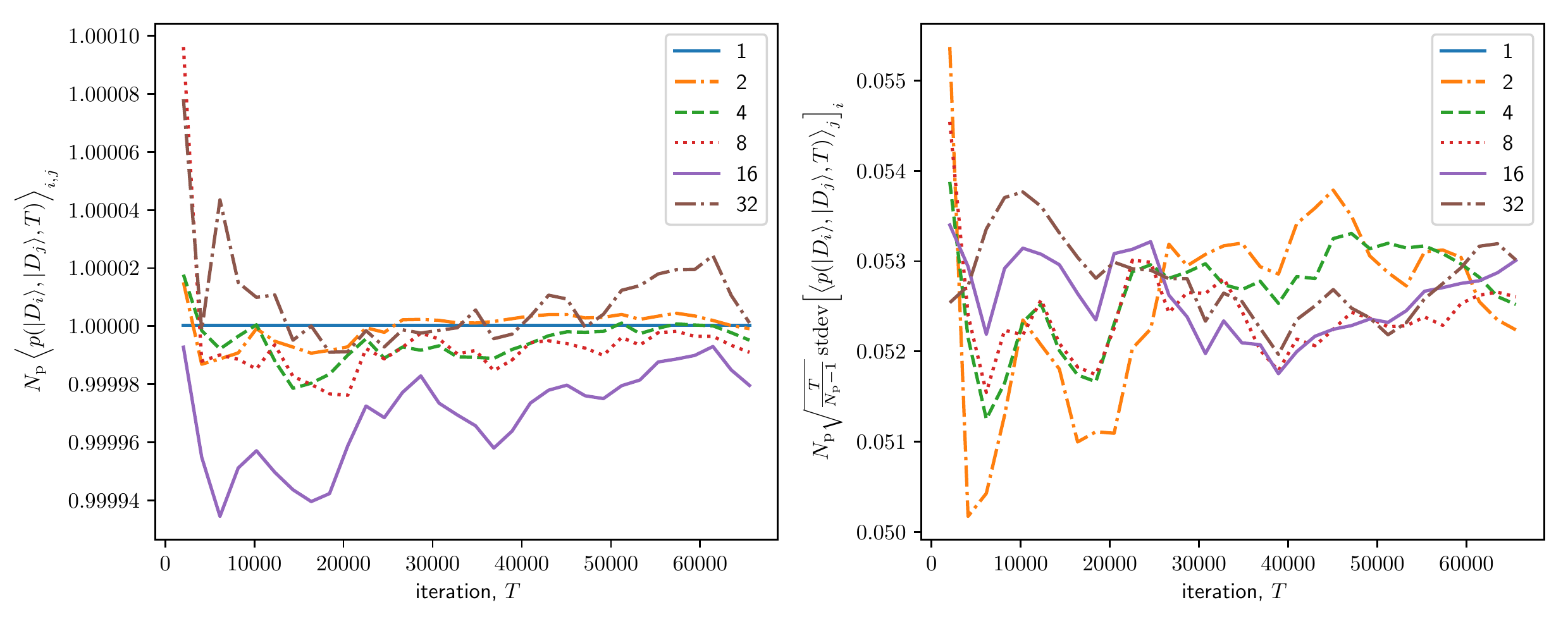}
        \caption{For $\nu_\mathrm{move}=4$, ({\it left}) the mean probability of two excitors sharing the same processor (rescaled by $1/{\Np^{-1}}$ such that the exactly correct value is unity); and ({\it right}) the standard deviation of the same probability (rescaled by $\Np\sqrt{\frac{T}{\Np-1}}$ to show the scaling behaviour).  The different lines correspond to different numbers of processors, $\Np$, given in the legend.}
    \end{subfigure}	
    \newline
	\begin{subfigure}[h]{\linewidth}
    \includegraphics[width=1.0\linewidth,keepaspectratio]{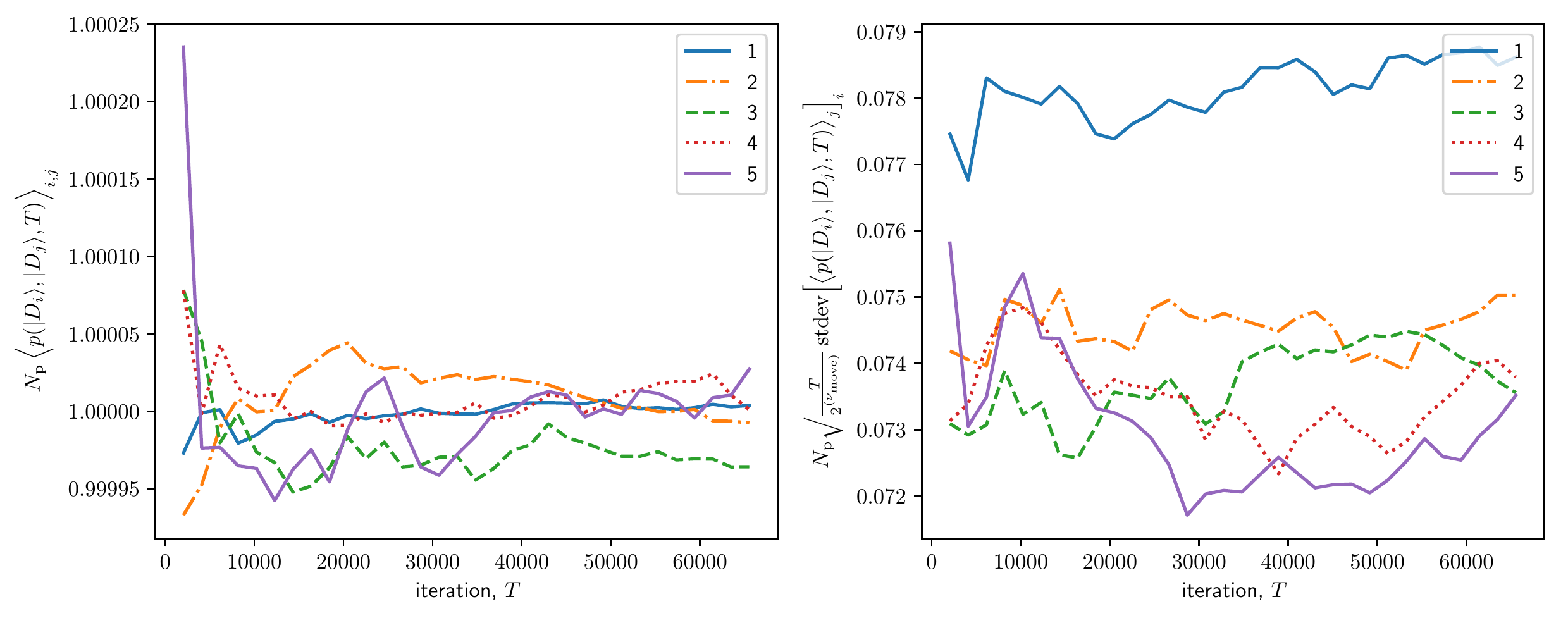}
        \caption{For $\Np=32$, ({\it left}) the mean probability of two excitors sharing the same processor (rescaled by $1/{\Np^{-1}}$ such that the exactly correct value is unity); and ({\it right}) the standard deviation of the same probability (rescaled by $\Np\sqrt{\frac{T}{2^{\Nmove}}}$ to show the scaling behaviour). The different lines correspond to different move frequencies, $\Nmove$, given in the legend. }
    \end{subfigure}	
    \caption{Analysis of the co-distribution of excitors for different $\Np$ and $\Nmove$ for a UEG system of 14 electrons in 186 spin-orbitals, consisting of 3813 excitors.  $p(|D_{\bm{i}}\rangle, |D_{\bm{j}}\rangle, T)$ denotes the probability that excitors on $|D_{\bm{i}}\rangle$ and $|D_{\bm{j}}\rangle$ were on the same processor, averaging over all iterations up to $T$.}
    \label{fig:excitdistrib}
\end{figure*}
The parallel CCMC algorithm can produce a biased estimate of the energy as not all
excitors can form clusters with \textit{all} available exitors in the spawn step and so
a subspace is sampled each iteration.
If the number of MPI processes is 1, then the complete CC space is sampled each iteration and
there is no bias.
Conversely if more than one MPI process is employed and $\Nmove = \infty$, then \cref{eq:ccmc_dist} reduces to \cref{eq:fciqmc_dist} and clearly the CC wavefunction cannot be sampled as excitors are fixed on specfic MPI processes and hence clusters involving excitors on different processes can never be sampled.
Whilst the dominant factor controlling the accessible subspace of clusters per iteration is the number of MPI processes, the timestep, $\delta\tau$, and (log of the) move frequency, $\Nmove$, are also important, as decreasing either amounts to increasing the available subspace per unit of imaginary time.

To demonstrate these effects, we have evaluated the correlation between the locations of excitors on a trial system of 3813 determinants.
The probability that a single excitor is on a specific processor may be regarded as a random event with probability $\frac1\Np$, and the long-time distribution of such events is expected therefore to follow a binomial distribution with this probability, giving an unbiased mean of $\frac1\Np$ and a variance which therefore scales with $\frac{\Np-1}{\Np^2}$.
Similarly, a move frequency of $\Nmove$ moves an excitor's processor every $2^{\Nmove}$ iterations, so decreases the variance by a factor of $2^{\Nmove}$.  
Both of these scaling effects translate directly to the correlated probability of two excitor locations. Figure \ref{fig:excitdistrib} shows no notable bias in the time-averaged mean probability of two excitors coinciding, and a standard deviation following the above scaling relationships.
Any bias present due to the instanteous probability distribution of excitors is can therefore be reduced by decreasing $\Np$ and $\Nmove$.

As a test system to see a bias, we consider the three-dimensional uniform electron gas (3D UEG)\cite{MartinUEGChapter, Giuliani2005, Loos2016a}, which
conveniently allows for an easily-adjustable Hilbert space and degree of correlation.
Specifically, we calculate the CCSDT energy of the 14-electron 3D UEG with 66 plane-wave
spin-orbitals at $\rs=0.5\bohr$ and $\rs=5\bohr$, for which parallelization unbiased results using solely OpenMP parallelization are
available\cite{Neufeld2017}. The full non-composite cluster selection algorithm (\cref{sec:fnc}) was used to aid convergence.
A discrepancy with magnitude of $0.01$eV/electron is similar in magnitude to chemical accuracy\cite{Foulkes2001,Wagner2016,Neufeld2017} and represents an upper bound on any bias, which would preferably be negligible.
Note that this is far from a production-level calculation: the CISDT Hilbert space for this system contains only 22969 determinants.

\begin{figure}
\centering
	\begin{subfigure}[h]{8.5cm}
    \centering
    \includegraphics[width=\linewidth,keepaspectratio]{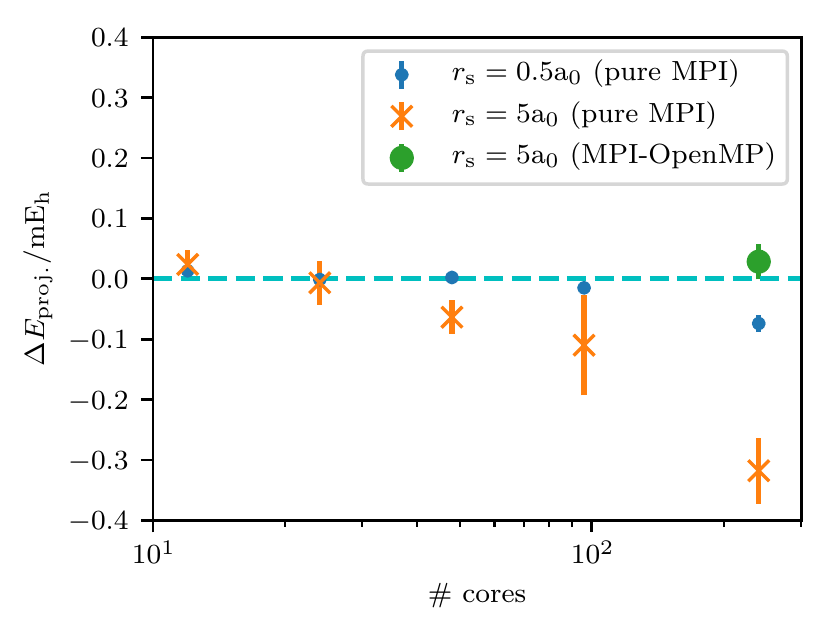}
        \caption{The bias in the CCSDT energy as a function of the number of cores using $\Nmove=5$ and $\delta\tau=0.001$. The MPI-OpenMP calculation used 12 OpenMP threads per MPI process.}
    \label{fig:MPIbias}
    \end{subfigure}	
    \begin{subfigure}[h]{8.5cm}
        \includegraphics[width=\linewidth,keepaspectratio]{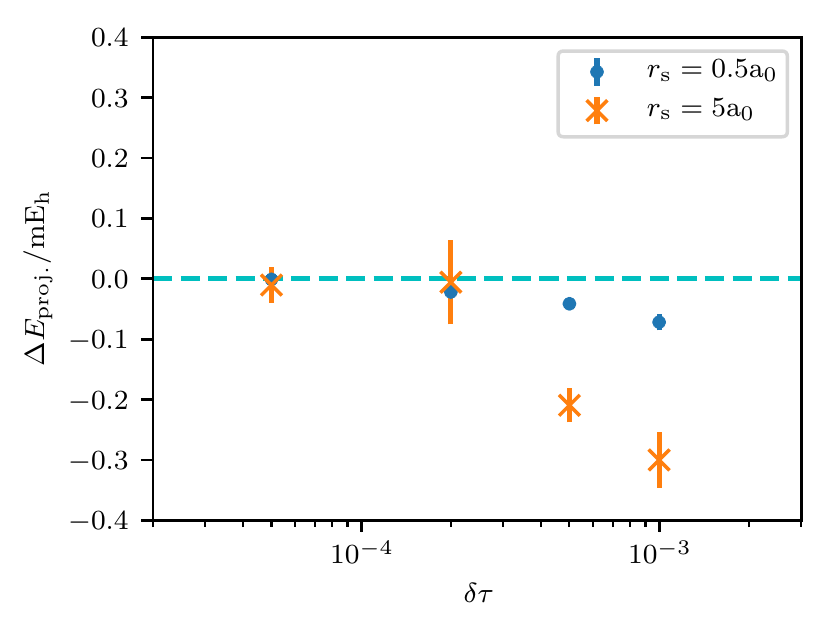}
        \caption{The bias in the CCSDT energy as a function of $\delta\tau$ using 240 MPI processes and $\Nmove=5$.}
	\label{fig:timebias}
    \end{subfigure}
    \newline
    \begin{subfigure}[h]{8.5cm}
	\includegraphics[width=\linewidth,keepaspectratio]{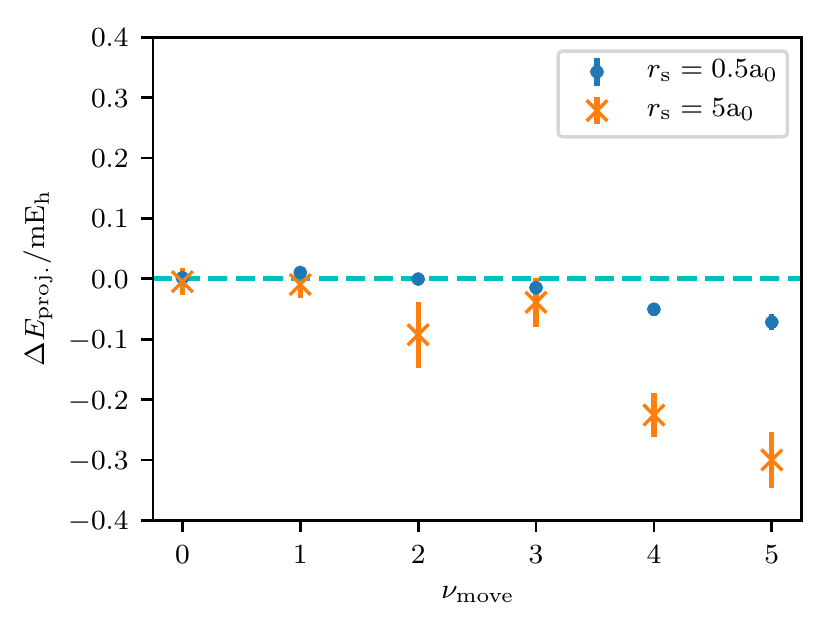}
        \caption{The bias in the CCSDT energy as a function of $\Nmove$ using 240 MPI processes and $\delta\tau=0.001$.}
	\label{fig:mfbias}
    \end{subfigure}
\vspace{-0.2cm}
    \caption{{The effects of the number of MPI processes, $\delta\tau$, $\Nmove$ on the deviation of the
        CCSDT projected energy from the unbiased value\cite{Neufeld2017} for the 14-electron 3D UEG with 66 spin-orbitals.
        An accuracy of $\pm 0.01$ eV/electron corresponds to 5 m\hartree here as we have 14 electrons. This is outside of the range shown. A horizontal line at
                zero error is shown to guide the eye.}}
    \label{fig:bias}
\end{figure}

\cref{fig:MPIbias} shows the dependence of the bias in the CCSDT projected energy as a function of the number of MPI processes for the UEG.
The bias increases with the number of MPI processes and is larger for $\rs=5\bohr$ than $\rs=0.5\bohr$.
At 240 MPI processes, each MPI process has fewer than 100 excitors (assuming perfect load balancing) and so the subspace spanned by each MPI process at any given timestep is very small. The degree of correlation is also important: 17\% of excips are on the reference for $\rs=0.5\bohr$ compared to just 2\% for $\rs=5\bohr$. As such the relative importance of products of clusters increases with correlation.
We have also performed a hybrid MPI-OpenMP calculation for $\rs=5\bohr$ using 20 MPI processes with 12 OpenMP threads per process for $\rs=5\bohr$ which agrees within error bars to the expected result and to the corresponding 20 MPI process calculation.

The bias can be reduced by decreasing the imaginary time a given subspace is sampled. \cref{fig:timebias,fig:mfbias} show reducing the timestep and move frequency respectively reduces the bias in the CCSDT energy using 240 MPI processes for the 3D 14-electron UEG system.
The bias remains smaller for the smaller values of $\rs$ value with otherwise identical parameters.
 
We wish to emphasise this is a contrived setup to show that it is possible to obtain biased results in extreme parameter ranges. 

Even ignoring computational and parallel efficiency, the small number of excitors per processor results in poor sampling of the cluster expansion and results in a biased sampling of the coupled cluster wavefunction.
We typically set the number of MPI processes such that each process contains at least $\mathcal{O}(10^5)$ excitors\footnote{We note that available computational resources rarely grow polynomially with system size!}. In addition to improved parallel efficiency, hybrid MPI-OpenMP parallelization greatly helps with this issue. As a demonstration of non-trivial calculations, we consider a larger UEG system and the water dimer.

The CISDTQ Hilbert space size of the 14-electron UEG in a basis of 358 spin-orbitals is 2.6(1)$\times10^8$.
Using 96 MPI processes (pure MPI parallelisation) gave an estimate of the CCSDTQ correlation energy at $r_s=1\,\bohr$ to be -0.51875(7)\,\hartree;
using 8 MPI processes each with 12 OpenMP threads (i.e. the same total resources) gave an estimate of -0.51866(7)\,\hartree.
A previous study exploiting only OpenMP parallelization found the correlation energy
to be -0.51856(7)\,\hartree\cite{Neufeld2017}.
The hybrid calculation agrees with the previous result within 2 standard errors (individual standard errors added in
quadrature) whereas the pure MPI calculation is close but does not agree within 2 standard errors.
However, while the Hilbert space is of the order of $3\times10^8$, the number of occupied excitors relevant to the calculation is only about $7\times10^6$--$8\times10^6$.
This means that in the pure MPI case, significantly less than $10^5$ excitors are on the same MPI process. We have used even selection \cite{Scott2017} for this CCSDTQ calculation.

The CCSDT energy of the \ce{H2O} dimer at its CCSDTQ optimized geometry obtained by Lane\cite{Lane_13JCTC} in the jun-cc-pVDZ basis set\cite{PapajakTruhlar_11JCTC} is compared to deterministic results calculated using MRCC\cite{MRCC,Rolik2013} in \cref{tab:prodbias}. The CCMC calculation employed the heat bath excitation generator\cite{Holmes2016a} with slight modifications\cite{Neufeld2018} and was run on 32 MPI processes each using 12 OpenMP threads.
The water dimer in the jun-cc-pVDZ basis has a Hilbert space of $1.16\times10^7$ at the CCSDT level. The stochastic wavefunction contained $\approx 7.9\times10^6$ excitors, resuling in $\approx 5.3\times10^4$ excitors per MPI process, assuming perfect load balancing. Both CCSDT results agree well with each other.
The CCMC CCSDT result is resolvably different from the CCSD and CCSDTQ energy within error bars, and no bias is visible.

\begin{table}
    \caption{Total energy $E_\mathrm{tot.}$ and the correlation energy $E_\mathrm{corr.}$ of the \ce{H2O} dimer in the jun-cc-pVDZ basis set using Hartree--Fock, deterministic coupled cluster from CCSD to CCSDTQ and CCSDT with CCMC using 32 MPI processes threaded into 12 OpenMP threads each. Even selection\cite{Scott2017} has been used. The units are hartrees.}
\begin{tabular}{lll}
Method  &  $E_\mathrm{tot.}$/\hartree &  $E_\mathrm{corr.}$/\hartree \\
\hline
Hartree--Fock & -152.0804195  &  0\\
CCSD deterministic  &  -152.5158272 &   -0.435407682\\
CCSDT deterministic  &  -152.5241192 &   -0.443699666\\
CCSDT QMC $12\times32$ cores & & -0.44369(7) \\
CCSDTQ deterministic & -152.5251164 &   -0.444696884\\
\end{tabular}
\label{tab:prodbias}
\end{table}

To demonstrate the capabilities and indicate the future possibilities of parallelized CCMC,
we have also studied a system of three water molecules separated at a large distance at the CCSDTQ level in a cc-pVDZ basis\cite{Dunning1989}. 
Using three molecules that are --- for all practical purposes --- infinitely separated has the advantage that
the total energy can be calculated by other means, as three times the energy of a single water molecule at CCSDTQ.
This is necessary as we have the deterministic calculation has proven too computationally expensive to be performed in a reasonable time.\footnote{Using MRCC\cite{MRCC}, a single iteration was not completed within a week on a 32-core node on the Cambridge Service for Data Driven Discovery (CSD3) operated by the University of Cambridge Research Computing Service (http://www.csd3.cam.ac.uk/).  This provides a lower estimate of 2 node-months, or 46000 core hours.}  

The calculation was restarted from a CCSDT QMC calculation and run with 400 MPI processes using
12 OpenMP threads each for part of the calculation that we are analysing. 
Even selection\cite{Scott2017} and the heat bath uniform singles\cite{Holmes2016a} excitation generator
\footnote{Terminology from Ref. \onlinecite{Neufeld2018} has been used. Single excitations were sampled uniformly,
double excitations with the heat bath excitation generator. Idea by and mentioned in Holmes et al. \cite{Holmes2016a}.} have been used.
In the CCMC calculation, the projected energy oscillates in the range -0.6507 to -0.6524 \hartree.
The true correlation energy, as found by using MRCC\cite{MRCC} scaling up
from one molecule, is -0.6515 \hartree which is included at around the middle of our range.
The CCSDT correlation energy, found by the same method, is -0.6501 \hartree, which is outside of the quoted
range for CCSDTQ with CCMC.
To give more than merely a range or to give a smaller, more certain, range, the calculation would have to be run for longer. However, the intent of this study is not to find a known CCSDTQ value
but to act as a demonstration that (parallelized) CCMC can give coupled cluster
energies that are not feasible with deterministic coupled cluster codes. Current developments of CCMC will enable more precise large calculations in the future.

\section{Improved stochastic sampling}
\label{sec:sampling}

The efficiency of a stochastic coupled cluster calculation is highly dependent upon the algorithm used to sample the various steps within the algorithm.
 The original implementation used a simple and easy-to-implement algorithm for selecting clusters from which to spawn. This has proved to be increasingly inefficient as system size increases. Scott and Thom have shown that an `even-selection' algorithm which selects clusters with probabilities more closely corresponding to their amplitude dramatically increases the stability of calculations\cite{Scott2017}.
In this section we describe some alternative approaches we have explored to improve the stochastic sampling,
\footnote{These approaches were conceived and implemented some time before the work in Ref.~\onlinecite{Scott2017}, and briefly referred to in the same, and are reported in fuller detail now for completeness.}
and first consider what metric we may use to measure this.
\begin{figure}
\includegraphics[width=1.0\linewidth,keepaspectratio]{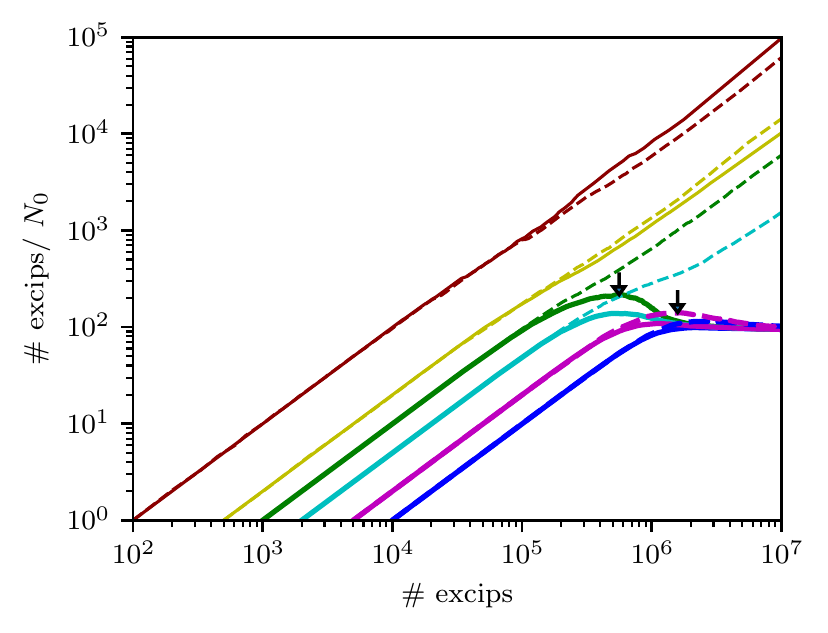}\\
\includegraphics[width=1.0\linewidth,keepaspectratio]{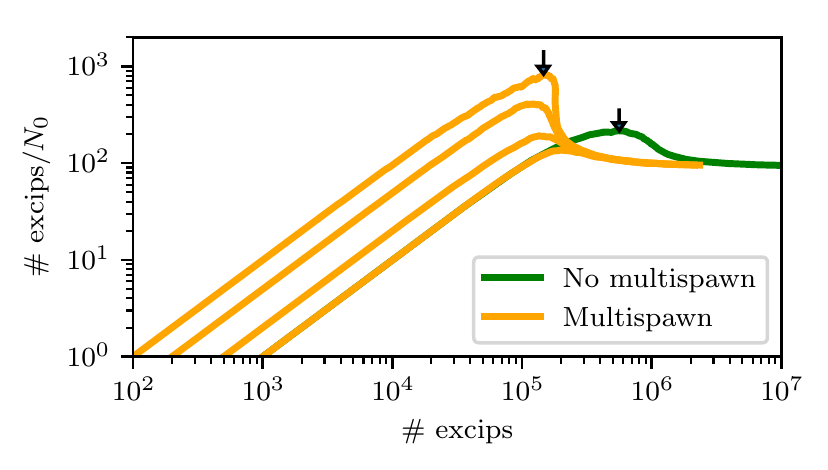}
\caption{Shoulder plots for frozen-core CCSD on benzene in a 6-31G basis using the original (dashed) and full non-composite (solid) algorithm. The initial population can be read from the intercept with the abscissa. ({\em top}) The arrows indicate the best estimate of the shoulder, showing that a full non-composite calculation has lower shoulder ($5.6\times10^5$) than the original algorithm ($1.6\times10^6$). ({\em bottom}) Multispawn, using $A_{\mathrm{thresh}}=1$, lowering the shoulder to $1.5\times10^5$, showing stability for many fewer initial particles.\\
Calculations used a timestep of $5\times10^{-4}$ and the renormalized excitation generators. $r_{\mathrm{CH}}=1.084$\,\AA\ and $r_{\mathrm{CC}}=1.397$\,\AA\ exploiting $D_{2h}$ symmetry.
}
\label{fig:fnc}
\end{figure}

\subsection{Shoulder heights}

The stability of a stochastic coupled cluster calculation can be determined by whether its population has overcome a plateau or shoulder in its dynamics\cite{Spencer2012}, where the rate of total particle growth (with imaginary time) slows or stops for some time while the correct  wavefunction is evolved.  A plateau is commonly visible in FCIQMC calculations whereas CCMC calculations typically only contain a shoulder in the total population growth. After this point, a calculation emerges with a stable growth rate of both the total and reference particle populations.  These factors are conveniently described on a shoulder plot\cite{Spencer2016}, which show a maximum in the ratio of total and reference populations at the shoulder.  The population at the shoulder is indicative of the relative difficulty of a calculation, and is affected by the parameters used to run it, and a best estimate is given with a low initial reference population and small timestep; using larger timesteps causes the shoulder population to increase.  As the population on the reference also dictates the normalization of the calculation, too low an initial population leads to unstable calculations which do not experience a shoulder and merely `blow up'.  Examples of shoulder plots are given in \cref{fig:fnc}.

In a stochastic coupled cluster calculation, the number of particles spawned at any step is directly proportional to the timestep, and after a certain threshold, larger timesteps will give a higher shoulder population.  To therefore determine the effects of any algorithmic changes we have used the parameters which give the lowest estimate of the shoulder, namely a very small timestep, and reducing the initial population, following a study in Ref. \onlinecite{Scott2017}
which compared selection algorithms with appropriate timesteps and initial populations.
\subsection{Non-composite clusters}
\label{sec:fnc}
We first look at reducing the noise due to stochastic sampling of the cluster expansion.
The total number of cluster selections to be made has previously been chosen to equal the total amplitude of excips, in analogy with FCIQMC, where each discrete psip individually undergoes spawning and death events.
In stochastic coupled cluster, clusters built up from single excips as well as multiple excips (known as composite clusters) are sampled, so the single-excip (non-composite) clusters have fewer samples taken than the total number of excips.
The sampling of (rather than explicit iteration through) the non-composite clusters proves to be an additional source of stochastic noise.
We therefore introduce a modification to the algorithm, {\em full non-composite} sampling, which explicitly iterates through the list of excips, performing spawning and death events on these individually.
It is still necessary to sample the composite clusters, and these are sampled with the same number of samples as total excips.
The effect of this sampling change is to reduce the number of particles at the shoulder, as shown in figure 
\ref{fig:fnc}, and so calculations require fewer excips to be stable. Even though the computational effort increases
since we are doing twice as many cluster selections, the number of minimum excips required 
is reduced by more than a factor of two
when using the full non-composite cluster algorithm as shown by the shoulder positions in the \textit{top} part of 
figure \ref{fig:fnc}. Figure 2 in Ref. \onlinecite{Scott2017} shows that at higher timesteps, the memory cost as a 
function of number of attempts per unit imaginary time is lower when using the full non-composite algorithm
compared to the original algorithm.
\subsection{Multiple spawning events}
\label{sec:multispawn}
The small timestep regime reduces the plateau height because at sufficiently small timestep all spawning attempts produce no more than one particle --- there are no `blooms', in which a single (rare) spawning event creates a large number of new particles on the same excitor.  Such rare events are undesirable both because of the inefficient exploration of the space (especially if the spawned particles do not have the same sign as the ground-state wavefunction) and the impact on population control. Unfortunately the computational efficiency of using such small timesteps is low, as the majority of spawning attempts produce no particles at all.  
Once a cluster has been selected and collapsed to produce determinant $\bm{j}$, the number of spawned particles depends on the amplitude of this cluster from which they are spawned.
As this is selected stochastically, the amplitude is unbiased by dividing by the probability of cluster selection to give an effective amplitude, $A(\bm{j})$.  The probability of spawning a particle from this cluster to determinant $\bm{i}$ is then given by 
\begin{equation}
p_{\mathrm{spawn}}(\bm{i}\leftarrow\bm{j})=\delta\tau\frac{|A(\bm{j})H_{\bm{ij}}|}{p_{\mathrm{gen}}(\bm{i}|\bm{j})}
\end{equation}
It is therefore possible to use the amplitude $A(\bm{j})$ to decide on the number of spawning attempts from that cluster.  A larger number of attempts to spawn will proportionately increase $p_{\mathrm{gen}}(\bm{i}|\bm{j})$, the probability that spawning onto $\bm{i}$ is attmepted from the excips on $\bm{j}$, and consequently decrease the spawning probability, reducing any blooms.  
The choice of when to change the number of spawning attempts should depend on $A(\bm{j})$, as merely keeping it at a constant, say $n$, would have a similar effect as using a timestep $\delta\tau/n$ and not account for the impact of rare events from clusters with large amplitudes.  We have chosen to introduce a threshold $A_{\mathrm{thresh}}$, such that the number of attempts is given by
\begin{equation}
n_{\mathrm{attempts}}=\max\left(1,\left\lfloor\frac{A(\bm{j})}{A_{\mathrm{thresh}}}\right\rfloor\right)
\end{equation}
The effect of these multiple spawning changes is shown in the lower panel of figure \ref{fig:fnc}, and in combination with the full non-composite sampling,  shows a significant reduction in shoulder heights and a corresponding decrease in memory requirements for calculations.
We finish with a note of caution when such sampling is used in combination with large-scale parallelization.
If the cluster sampling is such that there are occasional amplitudes significantly larger in magnitude than $A_{\mathrm{thresh}}$, a correspondingly large number of spawning attempts are made.
In some calculations $n_{\mathrm{attempts}}$ can be occasionally of the order of $10^6$.  Should such events be unevenly distributed over processors, it can lead to a significant load-imbalance and reduction of parallel efficiency.
In practice, we have only seen this effect when the number of excitors per process is under $10^5$.  In general such effects are averted by the use of even selection\cite{Scott2017} which ensures $|A(\bm{j})|$ takes an approximately constant value for all clusters. Even selection was used for the scaling plot in \cref{fig:scaling}.
\section{Discussion}
\label{sec:discussion}

Overall we have shown that, despite the non-linearity of the coupled cluster ansatz, by introducing a stochastic algorithm, it is possible to perform massively parallel calculations at arbitrary orders of coupled cluster theory with great parallel efficiency, and approximately ideal strong scaling up to 500 cores.
Our parallelization scheme exploits the stochastic nature of the algorithm to sample combinations of excitors averaged over multiple cycles, and we have shown that the bias introduced can be made minimal (provided the number of MPI processes and other parameters are chosen sensibly), well below the intrinsic accuracy of the calculations themselves.
Furthermore, the bias can be systematically reduced and so confidence can be had in its magnitude and as systems studied become larger, the parallelization bias becomes smaller.

We contrast our parallelization scheme to that of the Cyclops tensor framework\cite{Solomonik2014}, which requires explicit knowledge of the sparsity within the tensors 
of excitor amplitudes, and performs deterministic CCSD and CCSDT calculations.
While such a scheme produces numerically precise results, it cannot easily take advantage of the natural sparsity of amplitudes within excitation space, so requires significantly more storage for amplitudes.
While the polynomial scaling of the number of amplitudes with system size allows a good weak scaling behaviour, the requirement to communicate all  of these results in relatively poor strong scaling behaviour.
Efforts are underway\cite{JagodeDongarra2017_IJHPCA17} to redesign deterministic algorithms using a dataflow paradigm, however these require significant manual reorganisations of the code which appears infeasible for higher levels of coupled cluster theory.

We have also shown that the stochastic sampling can be improved using the \textit{full non-composite} or the \textit{multi-spawn}
additions. The later \textit{even selection} sampling\cite{Scott2017} was inspired by and was built on top of the
\textit{full non-composite} algorithm and when using multiple MPI processes is more efficient than \textit{multi-spawn}
 sampling which was a first step on the way to improving shoulder heights.

We close by noting that the approach described here is not restricted to just coupled cluster; rather the idea of sampling both the action of the Hamiltonian and the wavefunction ansatz is applicable to many other methods in quantum chemistry.

\begin{acknowledgments}
J.S.S.~acknowledges the research environment provided by the Thomas Young Centre under Grant No.~TYC-101.
V.A.N.~acknowledges the EPSRC Centre for Doctoral Training in Computational Methods for Materials Science for funding under grant number EP/L015552/1 and the Cambridge Philosophical Society for a studentship.
W.A.V.~ is grateful to EPSRC for a studentship.
R.S.T.F.~acknowledges CHESS for a studentship and A.J.W.T.~the Royal Society for a University Research Fellowship under grants UF110161 and UF160398.
This work used the ARCHER UK National Supercomputing Service (\url{http://www.archer.ac.uk}) under grant e507 and the UK Research Data Facility (\url{http://www.archer.ac.uk/documentation/rdf-guide}) under grant e507.
This work was also performed using resources provided by the Cambridge Service for Data Driven Discovery (CSD3) operated by the University of Cambridge Research Computing Service (\url{http://www.csd3.cam.ac.uk/}), provided by Dell EMC and Intel using Tier-2 funding from the Engineering and Physical Sciences Research Council (capital grant EP/P020259/1), and DiRAC funding from the Science and Technology Facilities Council (\url{www.dirac.ac.uk}).
\end{acknowledgments}

%
\end{document}